# Two-Dimensional Far-Field Correlations of X-ray Photon Pairs

E. Strizhevsky[1], Y. Klein[1,2], R. Hartmann[3], S. Francoual[4], T. Schulli[5], T. Zhou[6], A. Sharma[4], U. Pietsch[7], L. Strüder[3,7], D. Altamura[8], C. Giannini[8], M. Shokr[9], S. Shwartz[1*]

[1]*Physics Department and Institute of Nanotechnology, Bar-Ilan University, Ramat Gan, 5290002 Israel*
[2]*Nexus for Quantum Technologies, University of Ottawa, Ottawa, Ontario K1N 6N5, Canada.*
[3]*PNSensor GmbH, Otto-Hahn-Ring 6, Munchen, 81739 Germany*
[4]*Deutsches Elektronen-Synchrotron DESY, Notkestrasse 85, D-22607 Hamburg, Germany*
[5]*ESRF, The European Synchrotron, 71 Avenue des Martyrs, CS40220, 38043 Grenoble Cedex 9, France*
[6]*Center for Nanoscale Materials, Argonne National Laboratory, Lemont, IL, 60439, USA*
[7]*University of Siegen, Physics Department, Walter-Flex-Str. 3, Siegen, 57072 Germany*
[8] *Istituto di Cristallografia - Consiglio Nazionale delle Ricerche (IC−CNR), Via Amendola 122/O, I-70126 Bari, Italy*
[9] *European XFEL, 22869 Schenefeld, Germany*

We directly observe far-field correlations of x-ray photon pairs generated by spontaneous parametric down-conversion (SPDC). Using an energy-resolved, two-dimensional photon counting detector we record the full ring-shaped emission of both photons across a broad bandwidth and extract pair correlations directly from raw events without imposing angular constraints. The ring radii scale with photon energy, in quantitative agreement with transverse phase matching, providing a stringent momentum-space validation of x-ray SPDC. These observations open a route to leveraging quantum correlations in x-ray imaging and metrology, including correlation-enhanced magnification and reduced blurring.

Spontaneous parametric down-conversion (SPDC) generates correlated and entangled photon pairs and underpins a wide range of quantum-optical protocols, including heralded single-photon generation, entanglement-assisted measurements, and sub-shot-noise imaging [1–7]. Extending SPDC to the x-ray regime is particularly compelling: quantum correlations could enable imaging and metrology at ultralow fluence, with potential to radiation-sensitive samples and for fundamental tests of quantum optics at short wavelengths [8–15].

Despite substantial progress in demonstrating x-ray SPDC and related nonlinear effects [15–25], direct access to the joint transverse-momentum distribution of x-ray photon pairs has remained elusive. This limitation is not conceptual but experimental. The same crystal that produces SPDC pairs generates strong elastic and Compton-scattered backgrounds that exceed the pair rate by several orders of magnitude. Moreover, x-ray SPDC is intrinsically broadband in both angle and energy, rendering narrowband Bragg filtering ineffective for background rejection without simultaneously suppressing the biphoton flux. Consequently, experiments have typically relied on single-pixel, energy-resolving detectors and coincidence filtering [16–20], which restrict detection to a small angular fraction of the emission and require scanning apertures for imaging and correlation studies.

Pixelated detectors can, in principle, remove these constraints by enabling simultaneous acquisition over a large solid angle. However, in practice, their use has been limited by the need to identify true photon pairs in a high-flux environment, which requires simultaneous energy, temporal, and spatial resolution to discriminate correlated pairs from the large background of uncorrelated photons [26]. As a result, a direct experimental mapping of the two-dimensional joint transverse-momentum distribution of x-ray photon pairs has so far remained unavailable. Two recent reports of x-ray photon-pair observations using pixelated detectors achieved pair identification only by imposing angular-correlation constraints in the analysis [27,28].

In this Letter, we report a direct measurement of the two-dimensional far-field correlations of SPDC photon pairs across a broad spectral bandwidth without imposing angular-correlation constraints. This is achieved using a pixelated pnCCD detector [29,30] together with an event-based analysis procedure. We observe the ring-shaped far-field emission predicted by SPDC theory and directly resolve its energy dependence. In particular, the ring radii scale inversely with photon energy, in quantitative agreement with transverse phase matching, providing a stringent momentum-space validation of x-ray SPDC. Overall, this approach enables direct, assumption-free mapping of broadband x-ray biphoton correlations in momentum space and access to the full two-dimensional joint transverse-momentum distribution of x-ray SPDC pairs.

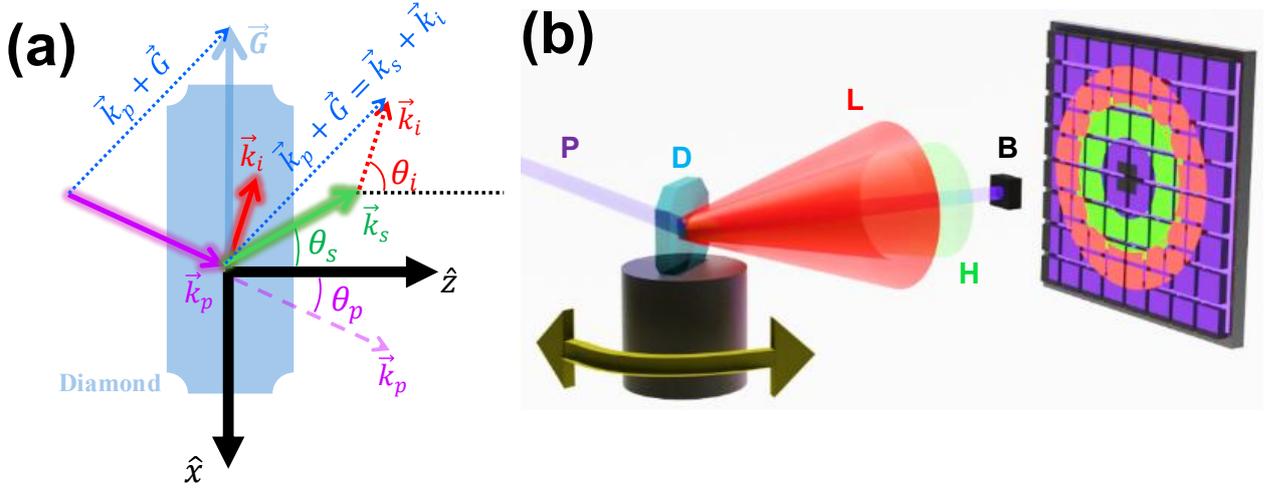

FIG 1(a). Phase matching condition diagram of the wave vectors $\vec{k}_p, \vec{k}_s$ and $\vec{k}_i$ of the pump, signal and idler photons respectively. $\vec{G}$ is the reciprocal lattice vector and $\theta_p$, $\theta_s$, and $\theta_i$ are the angles of the pump, signal, and idler wave vectors relative to the $\hat{z}$ direction, respectively. (b). Experimental setup: An x-ray pump beam (purple beam - P) is incident on a diamond single crystal in Laue geometry (D). The red and green cones represent the emission of lower (L) and higher (H) energy photons, respectively. A beamstop (B) is used to block the Bragg-diffracted beam.

X-ray photon pairs are generated through the interaction of an incident pump photon ($\hbar\omega_p$) with vacuum state fluctuations in a nonlinear medium. The pump photon undergoes spontaneous annihilation simultaneously producing two photons, signal and idler, with energies $\hbar\omega_s$ and $\hbar\omega_i$, respectively. Energy conservation requires $\hbar\omega_p = \hbar\omega_s + \hbar\omega_i$ and in the crystal the phase-matching requires $\vec{k}_p + \vec{G} = \vec{k}_s + \vec{k}_i$, where $\vec{G}$ is the reciprocal lattice vector, as shown in Fig. 1(a) [16,22].

The far-field correlations are governed by the phase-matching condition parallel to the crystal surface. Due to boundary conditions, this condition must be precisely satisfied along with the phase-matching condition in the direction normal to the surface, which can be nonzero. This results in a conical emission of the signal and idler radiation, centered around the vector $\vec{k}_p + \vec{G}$ [31]. Consequently, at the detector, we expect to observe rings centered around a point very close to the Bragg-diffracted signal. The ratio of the radii can be determined using the transverse phase-matching condition. For small emission angles measured relative to the Bragg angle and using n~1, it reduces to approximately:

(1) $$\frac{\hbar\omega_s}{\hbar\omega_i} = -\frac{\Delta\theta_i}{\Delta\theta_s}.$$

Here $\Delta\theta_s$, $\Delta\theta_i$ are the angular deviations of the signal and idler beams from the Bragg direction. Equation (1) therefore predicts that the ratio of the ring radii scales as the inverse of the photon-energy ratio, providing a direct momentum-space signature of biphoton correlations. Resolving this energy-dependent scaling requires simultaneous spatial and energy resolution and, to date, has not been observed in the x-ray regime. In the present work we leverage the pnCCD's combination of energy resolution, pixel size, and frame rate [29,30,32,33], together with an energy-gated, event-based analysis protocol, to directly measure Eq. (1) experimentally.

Our experiment was performed at the REXS first experimental station of beamline P09 at PETRA III synchrotron storage ring (DESY, Hamburg). The 21 keV incident beam was monochromatized using a Si (111) double-crystal monochromator [34]. The experimental setup is depicted in Fig. 1(b). To minimize background noise, primarily from Compton scattering, we positioned the detector at a 90-degree angle relative to the incoming beam [18–20]. The nonlinear medium was a diamond single crystal with dimensions of 4 mm x 4 mm x 0.8 mm with C(660) Bragg angle at 44.61°. The crystal was rotated 0.01° above the Bragg angle. To reduce accidental coincidences, we increased the detector frame rate to 1 kHz by x2 pixel binning along the vertical axis, resulting in 132x264 pixels with effective pixel size of 96x48 μm$^2$. We positioned the detector on the axis of the concentric emission cone at a distance of 200 mm from the nonlinear crystal. To prevent detector saturation, we narrowed down the upstream slits, reducing the photon flux to about $2\cdot10^{11}$ photons/s at the crystal.

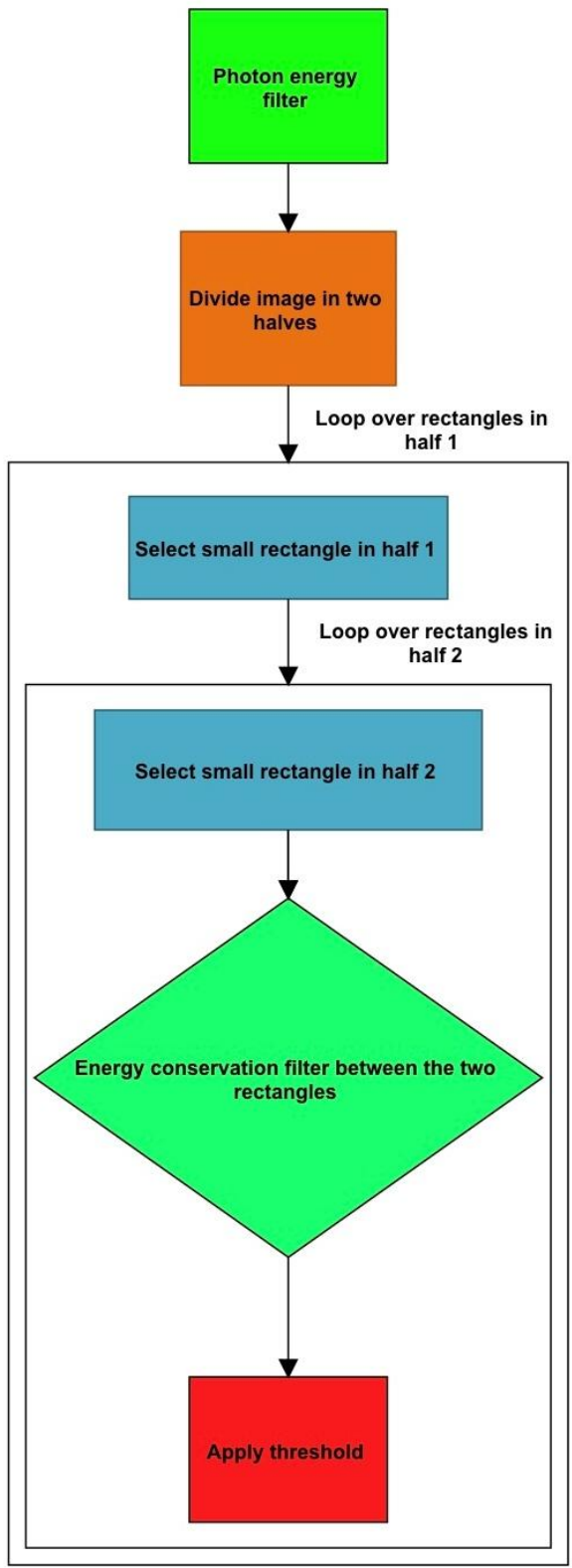

FIG. 2. Flowchart illustrating the algorithm used for extracting correlated photon pairs from raw experimental data: After initial photon-energy filtering, the detector image is divided into two halves. The algorithm then iteratively selects small rectangular regions from each half, applying an energy-conservation criterion between rectangle pairs. Events satisfying the energy criterion undergo a global thresholding step, identifying genuine photon-pair coincidences. See main text for further details.

We first preprocessed the pnCCD data by merging charge-sharing clusters into single-photon events, each tagged with position and photon energy. We masked defective pixels and grouped the remaining events into frames, each represented as a list of photon hits [31–33].

Despite the background-suppression geometry described above, each frame still contains multiple uncorrelated photon hits, so direct coincidence analysis is dominated by accidental events. We therefore isolate true pairs by enforcing energy conservation within small spatial subregions where the per-frame photon occupancy is sufficiently low for unambiguous pairing (Fig. 2). The detector is divided into two complementary half-planes about the Bragg-diffraction peak. Within these halves we define an 8 x 8-pixel region A (384 x 384 μm$^2$), and a 16 x 16-pixel region B (768 x 768 μm$^2$). The asymmetric region sizes compensate for the higher spatial density of states at higher photon energies, using smaller regions on the high-energy side and larger regions on the low-energy side.

A 400 eV energy window is applied in each region: region A retains photons with $\hbar\omega_s > \hbar\omega_p/2$, and region B retains photons with $\hbar\omega_i < \hbar\omega_p/2$. Candidate events are required to satisfy energy conservation within the bin widths. Regions A and B are then scanned across the detector and only events containing exactly one photon in A and one in B are retained. This procedure yields a two-photon hit array with high SNR across a broad range of energies. Correlated pairs are identified using a coincidence-count threshold of 20% of the maximum across all region pairs, chosen to provide a consistent contrast of $\cong 0.5$ across all analyzed energy bands. The true photon pairs rate is obtained by subtracting the mean false-coincidence background [31].

After applying the full selection procedure, for a representative pair of regions A and B centered at 10.5 keV, the residual accidental-pair rate is about 0.06 pairs per hour, whereas the true-coincidence rate is about 0.3 pairs per hour. This corresponds to a signal-to-background ratio of $\cong 5$. Importantly, the scan does not impose any a priori spatial correlation between A and B, preventing external bias. Applying the same pipeline to time-shuffled or events with randomized photon energies yields no structured correlation signal.

The procedure described above produces energy-dependent semicircular patterns on the detector, as shown in Fig. 3. The top row shows the experimental data, while the bottom row presents the simulation results.

The photon-energy windows used in Fig. 3, along with the corresponding coincidence count rates integrated over the full width at half maximum (FWHM) of the semicircular patterns, are summarized in Table I.

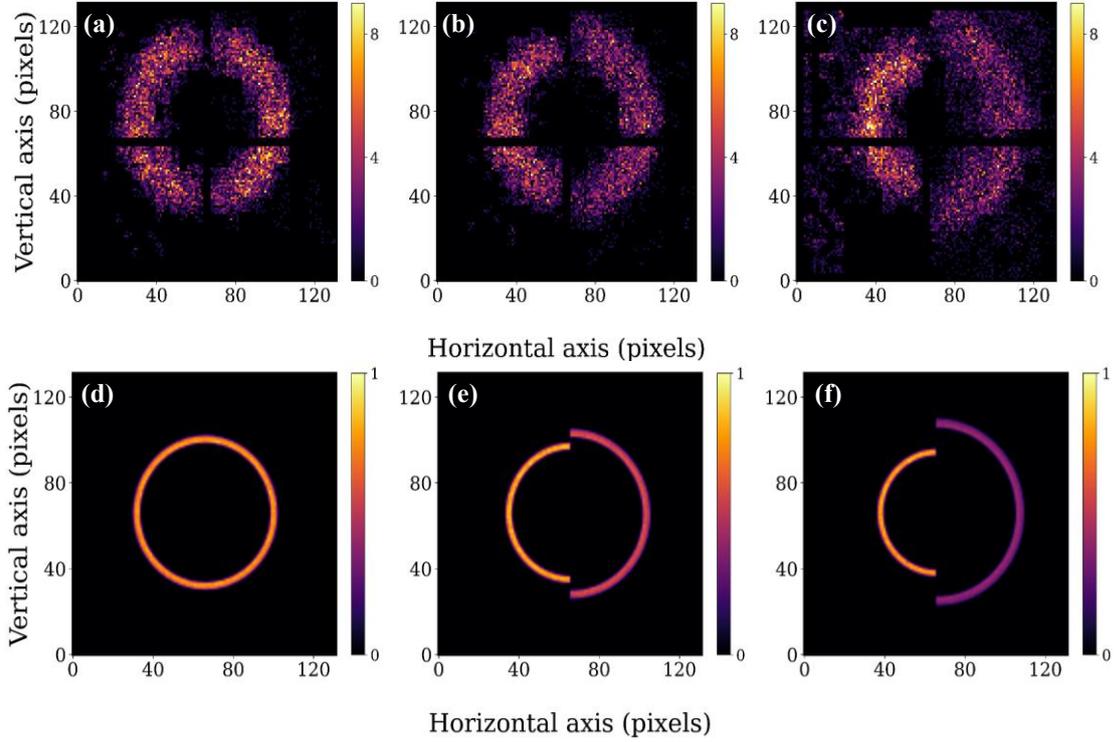

FIG. 3. – Experimental results compared with simulations. (a)–(c) Measured far-field distribution of photon pairs for the energy values listed in Table I. (d)–(f) The corresponding simulation results. Black vertical and horizontal stripes in (a)-(c), a few pixels wide and intersecting the ring region, arise from masked rows and columns of the detector and do not reflect physical features of the signal. The plotted data are shown with 2×2 pixel binning, corresponding to an effective pixel size of 96 x 96 μm².

| Panel | (a) and (d) | (b) and (e) | (c) and (f) |
|---|---|---|---|
| Left side energy window (keV) | 10.3-10.7 | 11.3-11.7 | 12.3-12.7 |
| Right side energy window (keV) | 10.3-10.7 | 9.3-9.7 | 8.3-8.7 |
| Coincidence rate (pairs per hour) | 124 | 97 | 69 |

TABLE I. Photon energy ranges used to select coincident photon pairs shown in the two halves of each correlation map and the corresponding total coincidence count rates per each energy window in Fig. 3.

The experimental correlation maps shown in Fig. 3 reveal clear energy-dependent semicircular features. These semicircular shapes are the direct far-field signature of pairwise transverse-momentum correlations imposed by the SPDC phase matching and boundary conditions. Their FWHM is 13-17 pixels, consistent with broadening dominated by the pump beam footprint, with an expected width of 15.5 pixels for a 0.5 mm beam and a 0.8 mm thick crystal. Additionally, the ring radii vary systematically with photon energy in quantitative agreement with theory. The total coincidence rate integrated over a bandwidth of 7 keV and an angular range of about 0.7 deg is $1260 \pm 35$ pairs per hour, in agreement with theory [35] and consistent with recent reports [27,28].

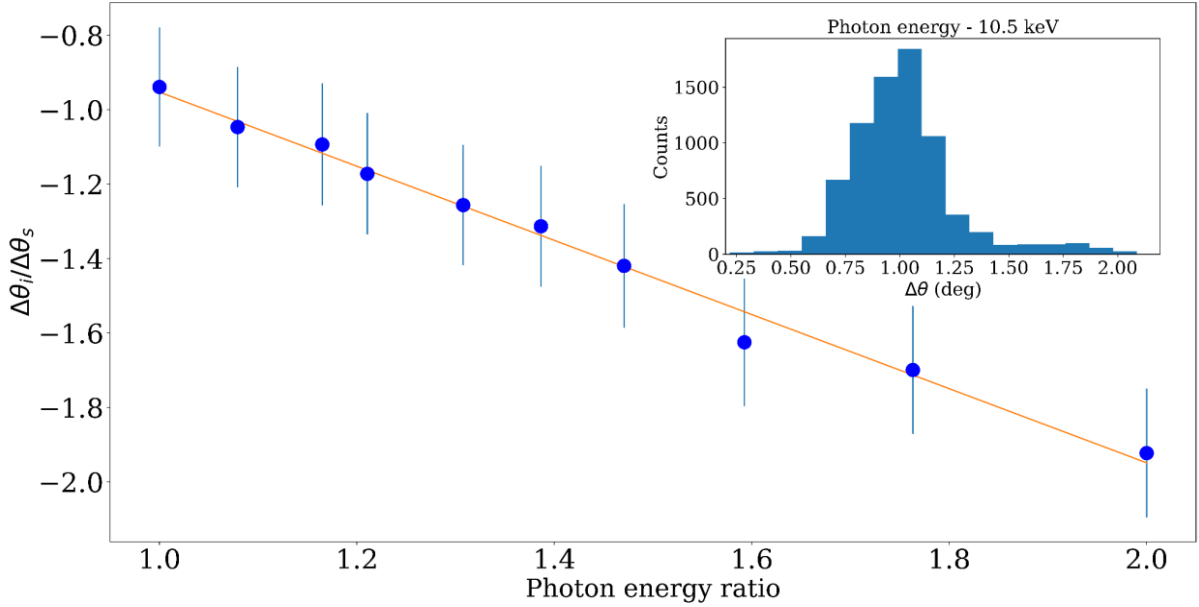

FIG. 4. Ratio of the signal and idler deviation angles as a function of the corresponding photon-energy ratio, $\omega_s/\omega_i$. Symbols show the experimental data, and the solid line is a linear fit. Each point is obtained from the centroid of a coincidence-rate histogram versus angular deviation from the Bragg angle for a given energy bin, as illustrated in the inset. The photon-energy bins have a bandwidth of 400 eV.

To quantify the energy dependence of the ring radii, we plot in Fig. 4 the ratio of the measured angular deviations, $\frac{\Delta\theta_i}{\Delta\theta_s}$, as a function of their corresponding photon-energy ratio $\frac{\hbar\omega_s}{\hbar\omega_i}$, together with the prediction of Eq. (1). The data exhibit the expected linear scaling, with a fitted slope of $-1 \pm 0.04$ in agreement with transverse phase matching for the present geometry and with numerical simulations. Verifying this scaling provides a stringent momentum-space test of x-ray SPDC and establishes a quantitative link between energy and transverse momentum in the biphoton far field.

Taken together, our measurements show that x-ray SPDC produces a well-defined biphoton far-field structure that can be exploited via its nonclassical correlations. One implication is quantum angular magnification: if the higher-energy photon interacts with the object, Eq. (1) enforces a correlated angular deviation in the lower-energy partner with a magnitude enhanced by the energy ratio, yielding tunable magnification upon detection of the partner photon. A second implication is quantum blurring reduction: detecting one photon near the source localizes the pair-creation point, and therefore the emission direction of its imaging partner becomes known, suppressing geometric blur [36]. Together, these effects point to correlation-enabled improvements in x-ray resolution and contrast beyond classical limits.

In conclusion, we have directly observed far-field transverse-momentum correlations of x-ray photon pairs generated by x-ray SPDC and resolved their energy-dependent ring structure. The measured radius scaling provides a quantitative, two-dimensional verification of energy-momentum relation in the x-ray regime. These results rely on a robust event-based analysis protocol that suppresses accidental coincidences and isolates true SPDC pairs over a broad bandwidth, without fast coincidence timing and without assuming any specific spatial correlation. This enables direct mapping of broadband x-ray biphoton correlations in momentum space. By establishing an experimental route to energy-resolved, two-dimensional biphoton correlations, our work positions x-ray SPDC as a controllable source of highly correlated x-ray photon pairs and enables correlation-based x-ray imaging and metrology at ultralow flux [12,37].

This work was supported by the Israel Science Foundation (ISF), Grant No. 2208/24. We acknowledge DESY (Hamburg, Germany), a member of the Helmholtz Association HGF, for the provision of experimental facilities. Parts of this research were carried out at beamline P09 of the PETRA III synchrotron light source and we would like to thank M. Sc. Julian Burkhardt for precise engineering and construction of the detector assembly on the two-theta arm of the psi diffractometer. Beamtime was allocated for proposal I-20211464 EC. Edward Strizhevsky thanks Dr. Andre Rothkirch for useful discussions regarding computing and the Maxwell resources. This research was supported in part through the Maxwell computational resources operated at DESY. We also acknowledge the European Synchrotron Radiation Facility (ESRF) for provision of synchrotron radiation facilities under proposal ID MI-1378 and on beamline ID01 [38].

References:

[1]   C. K. Hong, Z. Y. Ou, and L. Mandel, Measurement of subpicosecond time intervals between two photons by interference, Phys. Rev. Lett. **59**, 2044 (1987).

[2]   P. G. Kwiat, K. Mattle, H. Weinfurter, A. Zeilinger, A. V Sergienko, and Y. Shih, New High-Intensity Source of Polarization-Entangled Photon Pairs, Phys. Rev. Lett. **75**, 4337 (1995).

[3]   S. Friberg, C. K. Hong, and L. Mandel, Measurement of Time Delays in the Parametric Production of Photon Pairs, Phys. Rev. Lett. **54**, 2011 (1985).

[4]   G. Brida, M. Genovese, and I. Ruo Berchera, Experimental realization of sub-shot-noise quantum imaging, Nat. Photonics **4**, 227 (2010).

[5]   P. Kwiat, H. Weinfurter, T. Herzog, A. Zeilinger, and M. A. Kasevich, Interaction-Free Measurement, Phys. Rev. Lett. **74**, 4763 (1995).

[6]   S. Shwartz and S. E. Harris, Polarization Entangled Photons at X-Ray Energies, Phys. Rev. Lett. **106**, 80501 (2011).

[7]   M. O. Scully and M. S. Zubairy, *Quantum Optics* (Cambridge University Press, Cambridge, 1997).

[8]   N. Aslam, H. Zhou, E. K. Urbach, M. J. Turner, R. L. Walsworth, M. D. Lukin, and H. Park, Quantum sensors for biomedical applications, Nature Reviews Physics **5**, 157 (2023).

[9]   H. J. Kimble, The quantum internet, Nature **453**, 1023 (2008).

[10]  G. B. Lemos, V. Borish, G. D. Cole, S. Ramelow, R. Lapkiewicz, and A. Zeilinger, Quantum imaging with undetected photons, Nature **512**, 409 (2014).

[11]  A. G. White, J. R. Mitchell, O. Nairz, and P. G. Kwiat, Interaction-free imaging, Phys. Rev. A (Coll. Park). **58**, 605 (1998).

[12] H. Defienne, W. P. Bowen, M. Chekhova, G. B. Lemos, D. Oron, S. Ramelow, N. Treps, and D. Faccio, Advances in quantum imaging, Nat. Photonics **18**, 1024 (2024).

[13] A. Paniate, G. Massaro, A. Avella, A. Meda, F. V Pepe, M. Genovese, M. D'Angelo, and I. Ruo-Berchera, Light-field ghost imaging, Phys. Rev. Appl. **21**, 24032 (2024).

[14] A. Datta, Sensing with quantum light: a perspective, Nanophotonics **14**, 1993 (2025).

[15] R. Röhlsberger, J. Evers, and S. Shwartz, *Quantum and Nonlinear Optics with Hard X-Rays*, in *Synchrotron Light Sources and Free-Electron Lasers: Accelerator Physics, Instrumentation and Science Applications*, edited by E. J. Jaeschke, S. Khan, J. R. Schneider, and J. B. Hastings (Springer International Publishing, Cham, 2020), pp. 1399–1431.

[16] S. Shwartz, R. N. Coffee, J. M. Feldkamp, Y. Feng, J. B. Hastings, G. Y. Yin, and S. E. Harris, X-Ray Parametric Down-Conversion in the Langevin Regime, Phys. Rev. Lett. **109**, 13602 (2012).

[17] A. Schori, D. Borodin, K. Tamasaku, and S. Shwartz, Ghost imaging with paired x-ray photons, Phys. Rev. A **97**, 63804 (2018).

[18] D. Borodin, A. Schori, F. Zontone, and S. Shwartz, X-ray photon pairs with highly suppressed background, Phys. Rev. A (Coll. Park). **94**, 13843 (2016).

[19] E. Strizhevsky, D. Borodin, A. Schori, S. Francoual, R. Röhlsberger, and S. Shwartz, Efficient Interaction of Heralded X-Ray Photons with a Beam Splitter, Phys. Rev. Lett. **127**, 13603 (2021).

[20] S. Sofer, E. Strizhevsky, A. Schori, K. Tamasaku, and S. Shwartz, Quantum Enhanced X-ray Detection, Phys. Rev. X **9**, 31033 (2019).

[21] B. W. Adams, *Nonlinear Optics, Quantum Optics, and Ultrafast Phenomena with X-Rays: Physics with X-Ray Free-Electron Lasers* (Kluwer Academic Publisher, Norwell, MA, 2008).

[22] P. Eisenberger and S. L. McCall, X-Ray Parametric Conversion, Phys. Rev. Lett. **26**, 684 (1971).

[23] Y. Yoda, T. Suzuki, X.-W. Zhang, K. Hirano, and S. Kikuta, X-ray parametric scattering by a diamond crystal, J. Synchrotron Radiat. **5**, 980 (1998).

[24] N. J. Hartley et al., Confirming X-ray parametric down conversion by time–energy correlation, Results Phys. **57**, 107328 (2024).

[25] N. J. Hartley et al., *Down-Converted X-Ray Pair Generation at an X-Ray Free Electron Laser*, in *66th Annual Meeting of the APS Division of Plasma Physics* (APS, Atlanta, 2024), p. ZP12.00011.

[26] J. C. Goodrich et al., Quantum Imaging with X-rays, arXiv:2412.09833.

[27] N. J. Hartley et al., X-ray parametric down-conversion at an XFEL, Optica **12**, 961 (2025).

[28] J. C. Goodrich et al., Quantum correlation imaging via X-ray parametric down-conversion, Optica **13**, 135 (2026).

[29] S. Send, A. Abboud, R. Hartmann, M. Huth, W. Leitenberger, N. Pashniak, J. Schmidt, L. Strüder, and U. Pietsch, Characterization of a pnCCD for applications with synchrotron radiation, Nucl. Instrum. Methods Phys. Res. A **711**, 132 (2013).

[30] L. Strüder et al., Large-format, high-speed, X-ray pnCCDs combined with electron and ion imaging spectrometers in a multipurpose chamber for experiments at 4th generation light sources, Nucl. Instrum. Methods Phys. Res. A **614**, 483 (2010).

[31] See Supplemental Material at http:// for further details on the theoretical model, raw data analysis procedures, background estimation, and emission angle calculations.

[32] S. Ihle, P. Holl, D. Kalok, R. Hartmann, H. Ryll, D. Steigenhöfer, and L. Strüder, Direct measurement of the position accuracy for low energy X-ray photons with a pnCCD, Journal of Instrumentation **12**, P02005 (2017).

[33] R. Andritschke, G. Hartner, R. Hartmann, N. Meidinger, and L. Struder, *Data Analysis for Characterizing PNCCDS*, in *2008 IEEE Nuclear Science Symposium Conference Record* (2008), pp. 2166–2172.

[34] J. Strempfer, S. Francoual, D. Reuther, D. K. Shukla, A. Skaugen, H. Schulte-Schrepping, T. Kracht, and H. Franz, Resonant scattering and diffraction beamline P09 at PETRA III, J. Synchrotron Radiat. **20**, 541 (2013).

[35] Y. Klein, E. Strizhevsky, H. Aknin, M. Deutsch, E. Cohen, A. Pe'er, K. Tamasaku, T. Schulli, E. Karimi, and S. Shwartz, X-ray phase measurements by time-energy correlated photon pairs, Sci. Adv. **11**, eadw3893 (2026).

[36] A. Kueh, J. Warnett, G. Gibbons, J. Brettschneider, T. Nichols, M. Williams, and W. Kendall, Modelling the penumbra in Computed Tomography, J. Xray Sci. Technol. **24**, 583 (2016).

[37] M. Genovese, Real applications of quantum imaging, Journal of Optics **18**, 073002 (2016).

[38] S. J. Leake et al., The Nanodiffraction beamline ID01/ESRF: a microscope for imaging strain and structure, J. Synchrotron Radiat. **26**, 571 (2019).

# Supplemental material for: "Two-Dimensional Far Field Correlations of X-ray Photon Pairs"

E. Strizhevsky[1], Y. Klein[1,2], R. Hartmann[3], S. Francoual[4], T. Schulli[5], T. Zhou[6], A. Sharma[4], U. Pietsch[7], L. Strüder[3,7], D. Altamura[8], C. Giannini[8], M. Shokr[9], S. Shwartz[1*]

[1]Physics Department and Institute of Nanotechnology, Bar-Ilan University, Ramat Gan, 5290002 Israel
[2]Nexus for Quantum Technologies, University of Ottawa, Ottawa, Ontario K1N 6N5, Canada.
[3]PNSensor GmbH, Otto-Hahn-Ring 6, Munchen, 81739 Germany
[4]Deutsches Elektronen-Synchrotron DESY, Notkestrasse 85, D-22607 Hamburg, Germany
[5]ESRF, The European Synchrotron, 71 Avenue des Martyrs, CS40220, 38043 Grenoble Cedex 9, France
[6]Center for Nanoscale Materials, Argonne National Laboratory, Lemont, IL, 60439, USA
[7]University of Siegen, Physics Department, Walter-Flex-Str. 3, Siegen, 57072 Germany
[8] Istituto di Cristallografia - Consiglio Nazionale delle Ricerche (IC−CNR), Via Amendola 122/O, I-70126 Bari, Italy
[9] European XFEL, 22869 Schenefeld, Germany

## 1. Theoretical model

In this section, we provide detailed insights into our method for calculating photon pair rates and their two-dimensional far-field distributions. Our coordinate system is defined in Fig. S1. Essentially, the measured photon pair rates correspond to coincidence counts, reflecting simultaneous detection of signal and idler photons. To calculate the coincidence we use the second-order Glauber correlation function given by $R_c = S \iint \langle a_i^\dagger(\mathbf{r}_2, t_2) a_s^\dagger(\mathbf{r}_1, t_1) a_s(\mathbf{r}_1, t_1) a_i(\mathbf{r}_2, t_2) \rangle d\mathbf{u}\, d\tau$.

Here $S$ is the area of the pump at the input of the nonlinear crystal, $\mathbf{u} = \mathbf{r}_2 - \mathbf{r}_1$, and $\tau = t_2 - t_1$ [16]. Since calculations are typically simpler in the frequency domain than in space-time domain, we use the transformation: $a_j(z, \mathbf{r}, t) = \int_0^\infty \int_{-\infty}^\infty a_j(z, \mathbf{q}, \omega) \exp[-i(\mathbf{q} \cdot \mathbf{r} - \omega t)] d\mathbf{q} d\omega$, with $\mathbf{r} = (x, y)$ and $\mathbf{q} = (k_x, k_y)$. The angular frequency $\omega$ relates to the wavevector magnitude through $\omega_j = k_j c / n(\omega_j)$ and the wave-vector components $k_x$ and $k_y$ are parallel to the surfaces of the nonlinear crystal, as described in Fig. S1. The frequency domain operators satisfy the commutation relations: $[a_j(z_1, \mathbf{q}_1, \omega_1), a^\dagger_k(z_2, \mathbf{q}_2, \omega_2)] = \frac{1}{(2\pi)^3} \delta_{j,k} \delta(z_1 - z_2) \delta(\mathbf{q}_1 - \mathbf{q}_2) \delta(\omega_1 - \omega_2)$.

We first calculate the signal and idler operators at the nonlinear crystal output by solving the coupled equations in the Heisenberg picture under the undepleted pump approximation:

$$\frac{\partial a_s}{\partial z} = \kappa a_i^\dagger \exp(i\Delta k_z z)$$

$$\frac{\partial a_i^\dagger}{\partial z} = \kappa^* a_s \exp(-i\Delta k_z z) \tag{S1}$$

Here $\kappa$ is the coupling constant that includes the nonlinear coefficient and the pump intensity and $\Delta k_z$ is the phase mismatch along the z direction as described in Eq. S3. As previously shown, the coincidence count rate is obtained by numerical integration over photon energies within the relevant ranges [16]. Note that our model includes wavevectors with y-axis components, allowing us to simulate the two-dimensional far-field patterns of the emitted photon pairs. The simulated coincidence count rates plotted for angles $\theta$ and $\phi$ that satisfy Eq. S2, S3 and S4 result in the detailed two-dimensional patterns shown in Fig. 3 (d-f).

The phase mismatch in three dimensions is given by $\Delta \vec{k} = \vec{k}_p + \vec{G} - \vec{k}_s - \vec{k}_i$. The components of $\Delta \vec{k}$ along the axes are:

$$\Delta k_y = k_i \sin\phi_i - k_s \sin\phi_s \tag{S2}$$
$$\Delta k_z = k_p \cos\theta_p - k_s \cos\phi_s \cos\theta_s - k_i \cos\phi_i \cos\theta_i \tag{S3}$$
$$\Delta k_x = G - k_p \sin\theta_p - k_s \cos\phi_s \sin\theta_s - k_i \cos\phi_i \sin\theta_i \tag{S4}$$

Here, $k_p, k_s$, and $k_i$ denote the magnitudes of the pump, signal, and idler wave vectors, respectively, and $G$ is the magnitude of the reciprocal lattice vector, which is aligned along the x-axis. The coordinate axes and wave vector definitions are illustrated in Fig. S1.

Energy conservation, expressed as $\hbar\omega_p = \hbar\omega_s + \hbar\omega_i$, together with the boundary conditions enforcing exact phase matching in the plane parallel to the crystal surface ($\Delta k_y = \Delta k_x = 0$), enable the derivation of the approximate relationship between the emission angles of the signal and idler photons and their corresponding energies, as presented in Eq. 1 in the main text. For simplicity, the derivation was carried out in two dimensions, restricting the analysis to the plane defined by the pump direction and the reciprocal lattice vector $\vec{G}$. In this derivation, the refractive indices are approximated as unity, the trigonometric functions are expanded using a Taylor series around the Bragg angle (assuming $\theta_p, \theta_s$, and $\theta_i$ are close to this angle), and a total derivative is used to relate variations in the emission angles to changes in the photon energies.

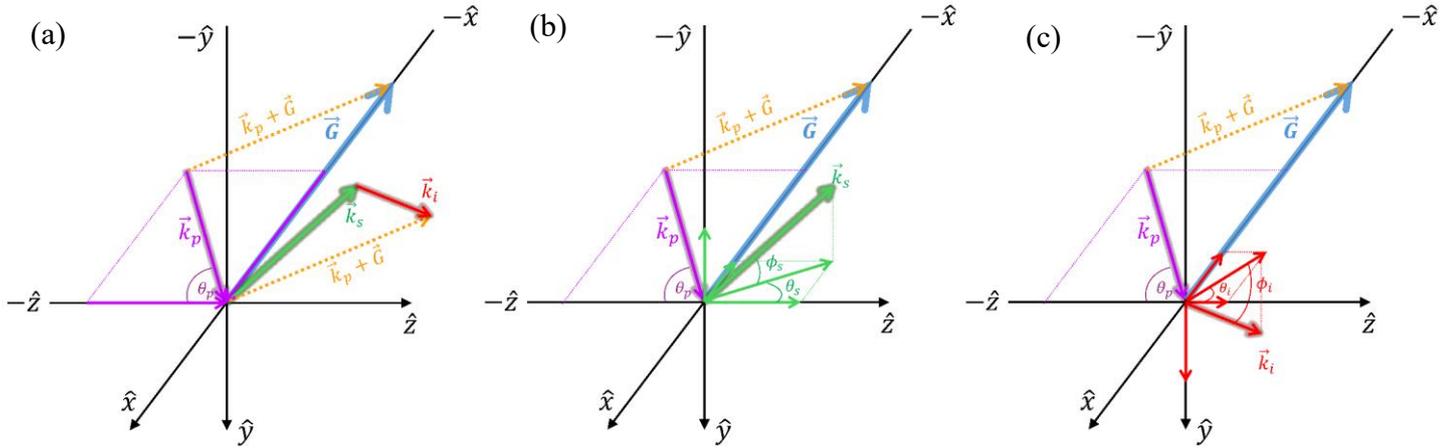

FIG. S1 (a) Three-dimensional representation of the phase-matching condition, showing the wave vectors $\vec{k}_p$, $\vec{k}_s$, and $\vec{k}_i$, with the signal $\vec{k}_s$ and idler $\vec{k}_i$ explicitly having components along the $y$-axis. (b) and (c) illustrate the decompositions of the signal $\vec{k}_s$ and idler $\vec{k}_i$ wave vectors into their respective coordinate components. The angles $\theta_p$, $\theta_s$ and $\theta_i$ denote the deviation angles of the pump, signal and idler wave vectors with respect to the z-axis in x-z plane, while angles $\phi_s$ and $\phi_i$ denote the deviation angles relative to the (x-z) plane (or the y-axis).

## 2. Procedure for Converting Raw pnCCD Frames to Photon Energies

Raw pnCCD frames were processed using standard gain and charge-transfer corrections [32,33]. A pixel gain map was applied to normalize the detector response and correct for pixel-to-pixel sensitivity variations. Column-dependent charge-transfer inefficiency corrections were then applied to compensate for charge losses during readout.

Photon events were reconstructed using cluster identification. Pixels exceeding the noise threshold were grouped into clusters corresponding to candidate photon events. Valid clusters were required to match predefined single-photon topologies consisting of 1-4 adjacent pixels. Clusters whose spatial distribution exceeded these patterns were rejected. To suppress artifacts from partial charge deposition or event overlap, pixels adjacent to rejected clusters were excluded from further analysis.

Additional pixel masks were applied to remove detector edge pixels, pixels adjacent to known defective pixels, and pixels outside the active detector region. The charge contained in each accepted cluster was summed and converted to photon energy using the detector calibration. An example of the resulting spectrum after applying this procedure is shown in Fig. S2, and a typical image corresponding to different energy ranges is presented in Fig. S3.

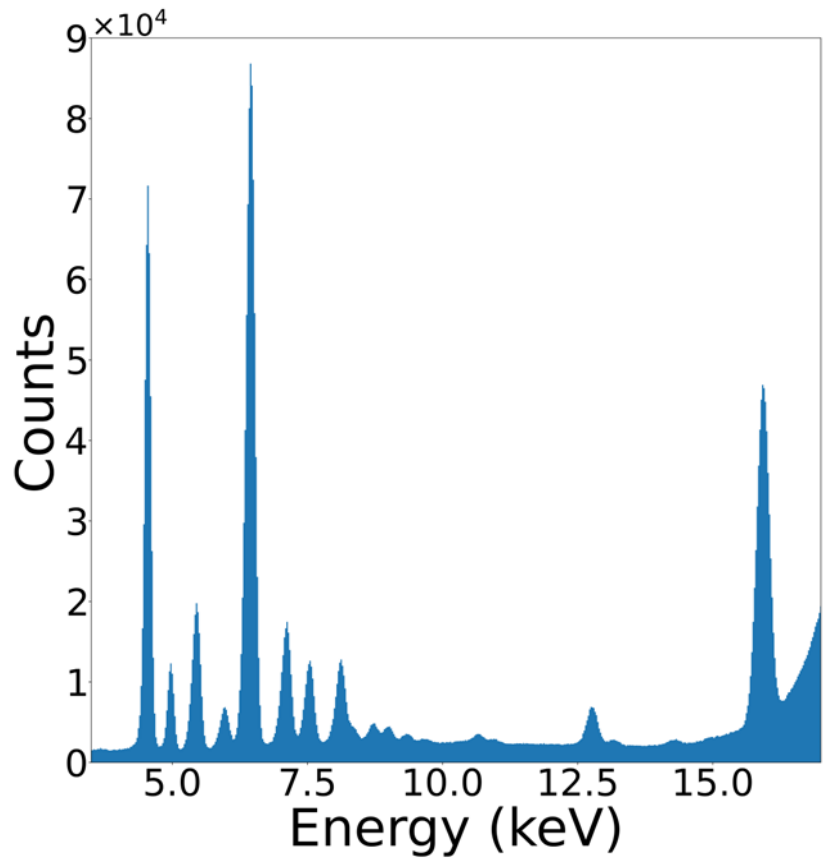

FIG. S2. Raw data spectrum for 600 seconds of measurement.

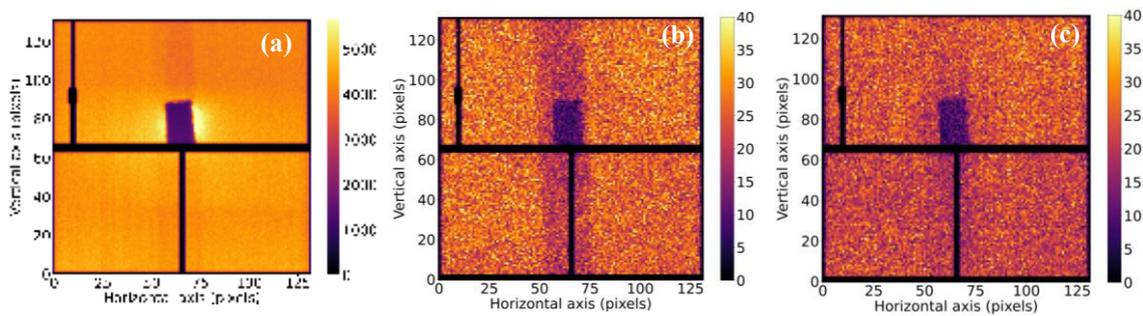

FIG. S3. – Detector data across different energy ranges before our analysis: (a) Full energy range, (b) 8.3–8.7 keV, and (c) 12.3–12.7 keV.

## 3. Background estimation

To quantify the rate of accidental coincidences arising from uncorrelated photons, we analyzed data in regions of the detector where no genuine signal was expected and only background photons were present. The data were collected over 31 hours. The number of coincidence events was evaluated as a function of photon energy, with particular attention to the energy windows where true photon-pair events are anticipated. This analysis provides a direct estimate of the accidental coincidence rate.

The elevated accidental coincidence rates observed in the final two entries of Table S1 are attributed to an increased background photon flux, consistent with the enhanced count rate in the photon spectrum below 7.5 keV shown in Fig. S2.

**Table SI.** Rate of accidental coincidence events within a 400 eV energy bandwidth, measured over regions of 8 × 8 pixels (for higher energy ranges) and 16 × 16 pixels (for lower energy ranges).

| Left side energy window (keV) | Right side energy window (keV) | Accidental coincidence |
|---|---|---|
| 10.3-10.7 | 10.3-10.7 | $0.042 \pm 0.002$ pairs/hour |
| 11.3-11.7 | 9.3-9.7 | $0.034 \pm 0.002$ pairs/hour |
| 12.3-12.7 | 8.3-8.7 | $0.059 \pm 0.003$ pairs/hour |
| 13.3-13.7 | 7.3-7.7 | $0.072 \pm 0.004$ pairs/hour |
| 13.8-14.2 | 6.8-7.2 | $0.01 \pm 0.006$ pairs/hour |

## 4. Emission Angles Calculation

The emission angle for each energy was obtained from histograms of photon counts as a function of distance from the ring center, which corresponds to the Bragg angle. This ring radius was defined as the average distance within the full width at half maximum (FWHM) of the histogram peak. The inset in Fig. 4 shows a representative histogram corresponding to the 10.5 keV data point (photon-energy ratio of 1). For this case, the average angular deviation from the Bragg angle is 0.94° with a FWHM of 0.4°. To improve the counting statistics the detector image were binned over 4 x 4 pixels. Histogram bins corresponded to two binned pixels (eight original pixels), improving the statistical reliability at the expense of spatial resolution. The uncertainty of each extracted radius was taken as one histogram bin above and below the calculated value. The measured radii were obtained in pixel units and converted to emission angles using the pixel size and the 200 mm detector distance from the diamond crystal.